\documentclass[11pt]{article}
\usepackage[margin=1in]{geometry}

\usepackage{sectsty}
\usepackage{graphicx}
\usepackage{amsmath}
\usepackage{amssymb}
\usepackage{amsthm}
\usepackage{float}

\newtheorem{definition}{Definition}

\newtheorem{innercustomgeneric}{\customgenericname}
\providecommand{\customgenericname}{}
\newcommand{\newcustomtheorem}[2]{%
  \newenvironment{#1}[1]
  {%
   \renewcommand\customgenericname{#2}%
   \renewcommand\theinnercustomgeneric{##1}%
   \innercustomgeneric
  }
  {\endinnercustomgeneric}
}

\newcustomtheorem{customlem}{Lemma}
\newcustomtheorem{customprop}{Proposition}

\usepackage[utf8]{inputenc} %
\usepackage[T1]{fontenc}    %
\usepackage{url}            %
\usepackage{booktabs}       %
\usepackage{amsfonts}       %
\usepackage{nicefrac}       %
\usepackage{microtype}      %
\usepackage{lipsum}
\usepackage{xcolor}

\usepackage{tabularx}
\usepackage{xcolor}
\usepackage{float,subcaption,placeins}
\usepackage{amsmath}
\usepackage{amsthm}
\usepackage{thm-restate}
\usepackage{float}
\usepackage{enumitem}
\usepackage{xr}
\usepackage{graphicx}
\usepackage{thmtools}
\usepackage{url}
\usepackage{bm}
\usepackage[normalem]{ulem}
\usepackage{balance}
\usepackage{lipsum}
\usepackage{etoc}
\usepackage{comment}
\usepackage{soul}
\usepackage{breakcites}

\usepackage[hyperindex,breaklinks]{hyperref}
\usepackage[capitalise,noabbrev]{cleveref}

\DeclareUrlCommand\email{\urlstyle{rm}}

\newcommand{\Comments}{1}
\newcommand{\mynote}[2]{\ifnum\Comments=1\textcolor{#1}{#2}\fi}

\newcommand{\parbold}[1]{\vspace{.25em}\noindent\textbf{#1}}

\usepackage[longnamesfirst]{natbib}
\renewcommand\cite[1]{\citep{#1}}

\newcommand\thetaset{\ensuremath{\{\theta_k\}_{k=1}^{K-1}}}

\newcommand\bbE{\ensuremath{\mathbb{E}}}
\newcommand\E{\ensuremath{\mathbb{E}}}

\usepackage[group-separator={,}]{siunitx}
\newcommand\blfootnote[1]{%
	\begingroup
	\renewcommand\thefootnote{}\footnote{#1}%
	\addtocounter{footnote}{-1}%
	\endgroup
}

\usepackage{pgfplots}

\pgfmathdeclarefunction{gauss}{2}{%
  \pgfmathparse{1/(#2*sqrt(2*pi))*exp(-((x-#1)^2)/(2*#2^2))}%
}

\usetikzlibrary{patterns}

\pgfplotsset{compat=1.10}
\usepgfplotslibrary{fillbetween}

\usepackage{xcolor}
\definecolor{babyblueeyes}{rgb}{0.63, 0.79, 0.95}
\definecolor{powderblue}{rgb}{0.69, 0.88, 0.9}

\usepackage{chngcntr}
\usepackage{apptools}

\newcommand\eqdist{\stackrel{\mathcal{D}}{=\joinrel=}}

\newcommand\neqdist{\stackrel{\mathcal{D}}{=\joinrel\neq\joinrel=}}

\usepackage{authblk}

\title{Test-optional Policies: Overcoming Strategic Behavior and Informational Gaps}

\author[1]{Zhi Liu}
\author[1,2]{Nikhil Garg}
\affil[1]{Operations Research and Information Engineering, Cornell University}
\affil[2]{Cornell Tech and Technion}
\date{\today}

\begin{document}
\maketitle

\begin{abstract}

Due to the Covid-19 pandemic, more than 500 US-based colleges and universities went ``test-optional'' for admissions and promised that they would not penalize applicants for not submitting test scores, part of a longer trend to rethink the role of testing in college admissions. However, it remains unclear how (and whether) a college can simultaneously use test scores for those who submit them, while not penalizing those who do not--and what that promise even means. We formalize these questions, and study how a college can overcome two challenges with optional testing: \textit{strategic applicants}  (when those with low test scores can pretend to not have taken the test), and \textit{informational gaps} (it has more information on those who submit a test score than those who do not). We find that colleges can indeed do so, if and only if they are able to use information on who has test access and are willing to randomize admissions.\blfootnote{\email{zl724@cornell.edu}, \email{ngarg@cornell.edu}}\blfootnote{The authors are grateful to Alex Wei for insightful discussions at the outset. We also thank Hannah Li, Faidra Monachou and Wenchang Zhu for helpful comments.}

\end{abstract}

\section{Introduction}

\begin{quote}
    By going test-optional, institutions are making a definitive statement that they will not need test scores to make admission decisions this year. Despite the change in policies, high school students and their parents are asking, ``Does test-optional really mean test-optional?'' The answer, simply put, is: YES. The following colleges with test-optional policies in place affirm that they will not penalize students for the absence of a standardized test score.~\cite{testoptional21}
\end{quote}
Due to the Covid-19 pandemic, more than 500 US-based colleges and university systems -- including the University of Texas at Austin, many other state universities, and Harvard -- adopted ``test-optional'' policies for college admissions and signed the above pledge; students do not have to submit standardized test scores  (the SAT or ACT) as part of their application. While mass adoption of test-optional policies---motivated by testing center closures and other pandemic-related academic disruptions---may not last far beyond the 2020-2021 admissions cycle, the push comes amid a growing movement to reconsider the role of standardized testing in college admissions. Independent of the pandemic, the entire University of California (UC) system recently settled a lawsuit by agreeing to no longer consider SAT or ACT scores altogether, even for students who submit them~\cite{delrio_2021}; this decision overrode a previous decision to go test-optional for Californian applicants and to eventually design their own test~\cite{watanabe_2020}. Test-optional policies further have a long history in college admissions in the United States~\cite{buckley2018measuring}. %

Why do schools consider test-optional policies, as opposed to either requiring scores for all applicants or not considering scores at all? Naively, such policies capture the benefits of testing without its downsides. Test supporters assert that tests simply provide additional information, the proper use of which should improve decision-making~\cite{phelps2005defending}. This additional information may especially help applicants from non-traditional backgrounds; a UC report on testing found that ``consideration of test scores allows greater precision when selecting from [under-represented minority] populations''~\cite{berkeleyreport}, and a New York Times article posited that without tests, ``a capable student from a little-known school in the South Bronx may be more challenging to evaluate''~\cite{bellafante_2020}. On the other hand, critics assert that test requirements may prevent some students from applying altogether, if they could not take the test. Over half the students who registered to take the SAT in September, 2020, had their registrations canceled due to testing center closures~\cite{testcentersclosed20}. Even in regular years, there is disparate access to testing:
``for every ten poor students who score college-ready on the ACT or SAT, there are an additional five poor students who would score college-ready but who take neither exam''~\cite{hyman2016act}; students outside the US may have to travel to different countries to take the test~\cite{internationalrtavel19}.  See~\citet{garg2020standardized}, which studies the role of these competing effects on the decision to drop testing altogether, for an extended discussion.\footnote{There are other benefits (such as biases in other features) and criticisms (like culturally biased nature of tests). We focus on the informational and access aspects.}

Test-optional policies seem to capture the best of both worlds, enabling some students to distinguish themselves with a high score, while allowing students without access to apply. However, despite the strong language in the pledge to ``not penalize students for the absence of a standardized test score,''~\cite{testoptional21}, it is unclear how colleges can implement test-optional policies when students are competing for limited spots. In the extreme case, consider two students exactly the same in every respect, except that one also has a (high) standardized test score and the other does not submit a score. It seems as if the school must either ignore the test score, or implicitly ``penalize'' the student without it. In this work, we ask: \textit{What can ``test-optional'' even mean, and how (and under what conditions) can a school use the test score without ``penaliz[ing]'' students without access?}

To answer this question, we model ways in which a school may implement a ``test-optional'' policy, and the resulting implications for various fairness notions and for student strategies. We find a stark divide between two settings: when the school knows (or can accurately estimate) whether a student had access to the test (whether or not they take it), and when it does not. When the school knows who has access, it can achieve a ``moderate'' fairness guarantee, albeit by following a non-Bayesian optimal admissions scheme (that as a result ranks some individuals ``inaccurately''). This scheme is akin to Thompson sampling; it simulates each students' potential test score if they do not report one; a school in practice that used such a scheme would admit some students who may otherwise be below the admissions cut-off---if the school believes that a test score could have helped them.   
On the other hand, when the school does not know and cannot estimate test access, being fair to those without test access becomes equivalent to ignoring test scores for everyone altogether---in our model, not even mandating reporting of scores conditional on taking the test would help counter-act the implications of strategic behavior.%

This divide stems from the differences between the two primary implications of optional testing amid differential access. First, there is a \textbf{strategic} component: those with access can choose to either not take the test or not report their score if it is to their benefit. If the school does not know who has access, we find that any attempt to be fair to those without access would be undone by students with access acting as if they do not. Second, there is an \textbf{informational} effect: even if students do not strategically with-hold their scores, the college has more data on students who take the test, and so potentially more evidence that they may be ``above the bar'' for admissions. In our model, a school with distributional knowledge of the parameters can counter-act the information gap by following the non-Bayesian scheme outlined above {across the population}.%

Our work has policy implications for schools navigating admissions under test-optional policies: they cannot be naive in their implementation. First, they must  target their interventions to those truly without test access, such as by using contextual information from elsewhere in a student's application (e.g., socioeconomic information or how many students from their high school took the test) or asking students to credibly attest to a lack of access. Second, they must be willing to take a chance on some students without a test score, if given a high enough test score their admissions outcome would have changed. %

The paper is organized as follows. Next, we discuss related work. \Cref{sec:model} contains our model. \Cref{sec:schooldoesnotknow} considers the case when the school does not know who has test access; as our fairness notions are impossible in such a setting, in \Cref{sec:schoolknows} we turn to the setting when the school knows who has access. We conclude in \Cref{sec:conclusion}. Proofs are deferred to the appendix.

\subsection*{Related work}

Our work broadly relates to the fair machine learning/mechanisms literature related to admissions and statistical discrimination in economics. It further contributes to the (primarily empirical) literature on optional testing. 

\parbold{Education and allocation in Fair ML} 
Early works studying fairness in education traditionally focused on problems such as school choice and affirmative action in college admissions, e.g. \citet{abdulkadirouglu2003school} and \citet{foster1992economic}. Our work is more closely connected to recent work in both the machine learning and mechanism design communities on fairness in ranking and allocation in general, and on education in particular~\cite{cai_fair20, emelianov2020fair, faenza2020impact, kannan2019downstream, haghtalab2020maximizing, immorlica2019access, Kleinberg_Mullainathan_2019, liu2020disparate, mouzannar2019fair}. For a review of fairness in both machine learning and mechanism design, see \citet{finocchiaro2021bridging}.

Our work extends that of~\citet{garg2020standardized}, which develops a multiple-feature model to analyze the decision to drop testing altogether---when two socioeconomic groups differ in how informative application components are about them and the proportion of students in the group who have access to the standardized test. We rather analyze test-{optional} policies, which entails considering students' strategic responses; motivated by the colleges' above pledge, we consider fairness notions for applicants without access, as opposed to for (observed) socioeconomic groups. 

Most related are the works of \citet{Kannan_Niu_Roth_Vohra_2021} and \citet{krishnaswamy2020classification}. \citet{Kannan_Niu_Roth_Vohra_2021} consider a setting in which some students can take a (binary) standardized test multiple times, while other students can only take it once. They then analyze two reporting policies: whether students must report all their scores, or just their best score. \citet{krishnaswamy2020classification} consider the computation of classifiers robust to strategic reporting, in a setting where the ``revelation principle'' holds: there always exists a strategy-proof classifier that performs as well as the best non-strategy-proof classifier. Our model and research questions differ in several key respects from both works. Most substantively, we consider a setting in which test scores are just one of multiple application components, and so students may strategize on whether to even \textit{take} the test; 
further, we are primarily motivated by the schools' pledge regarding optional tests, and the design of mechanisms that can meet such a pledge.

Finally, as in the work on strategic classification and its disparate impacts~\cite{Hardt:2016:SC,hu2019disparate,milli2019social,haghtalab2020maximizing,alon2020multiagent,braverman2020role}, we analyze how the strategic reactions of (some) applicants to a learning mechanism differentially impacts applicants.

\parbold{Economics of discrimination and measurement} This work proposes a model that draws upon the theory of \textit{statistical discrimination} introduced by \cite{phelps1972statistical} and \cite{arrow1971}. Such an approach has been used recently to study admissions \cite{emelianov2020fair,Temnyalov_2018,garg2020standardized}. Our analysis further exhibits the effects of \textit{adverse selection} \cite{akerlof1970market}, where informational asymmetries create an imbalance between the sellers and the buyers that may lead to a market collapse. In our model, the informational asymmetry between the school and the applicants in many cases induces students to conceal their test score. Though there are an abundance of works on adverse selection in economics, particularly the insurance literature, e.g. \cite{wilson1977model,dionne1992adverse}, its effects in the college admission process has not been seen studied theoretically; we illustrate how to design admissions policies to avoid such adverse selection and when it is unavoidable.

Finally, like us,~\citet{li2020hiring} propose a randomized (interview) policy in the presence of uncertainty in an applicant's potential, finding empirically it improves both quality and demographic diversity of candidates.

\parbold{Optional testing} This work also draws from empirical studies of optional testing from both college practitioners and researchers, where evidence of strategic behavior and unfairness can be observed. Test-optional admission in the United States has a long history; in 1969, Bowdoin College went test-optional, to focus admission to Bowdoin ``on the human quality of its students'' \cite{belasco2016enlightened}. Only in recent decades have researchers studied the effects of such policies, however.  After examination of a large-scale sample of institutional data collected over an eight-year period, \citet{hissfranks2014} find a bi-modal income distribution among non-submitters, where both students from both underprivileged and privileged background were prominent. \citet{belascoetal2014} report that on average, institutions that went test-optional saw a 26\% increase on reported SAT scores.\footnote{This difference in reported scores may actually be preferred by colleges, since it helps them manipulate their college ranking with higher average \textit{reported} SAT scores among admitted students \cite{ehrenberg2005method}. We do not consider such alternative college objectives, and instead take them at their word regarding satisfying their aforementioned pledge.} \citet{morgan2016} and \citet{robinsonmonks2005} find that non-submitters, if admitted, were more likely to enroll than submitters. This evidence suggests a strategic component of score reporting. More directly, \citet{wainer2011} finds that among the admitted students of Bowdoin College under a test-optional policy, students who elected not to submit their SAT scores earned a score on average 120 points lower than those who submitted, a difference of about one standard deviation. See \citet{buckley2018measuring} for an extensive discussion on the empirical literature on optional testing.
These empirical results, if repeated, would contradict the NACAC pledge to not penalize students without a test score: if students who have access strategize whether to report their score, they benefit at the cost of students who cannot take the test in the first place. %

\section{Model}
\label{sec:model}

Our model extends that of~\citet{garg2020standardized}, such that a test score can be optionally submitted. %

\parbold{Students. } There is a unit mass of students. Each student has a latent skill level $q$, normally distributed as $\mathcal{N}(\mu,\sigma^2)$, as well as a set of observed features $\Theta$. All students have the first $K-1$ features $\{1,\dots,{K-1}\}$; only a subset apply with the last feature $K$. For each student, $\theta_k$ is a noisy function of $q$, i.e., $\theta_k = q + \epsilon_k$, for $k = 1,\dots,K$, with Gaussian noise $\epsilon_k\sim \mathcal{N}(0,\sigma_k^2)$. The distribution of noise $\epsilon_k$ is feature dependent, but independent across features and students. Throughout our exposition, we refer to feature $K$ as the ``test score.'' 

For each student, let $Z, Y, X \in \{0, 1\}$ be indicators denoting whether the student has \textit{access} ($Z$) to the test, \textit{takes} ($Y$) it, and \textit{reports} their score ($X$), respectively. Of course, $Z \geq Y \geq X$. Access $Z$ is deterministic and pre-set, uncorrelated with $q$ and other features. If a student has access ($Z=1$), how they make their decisions to take the test and/or report is defined below.

\parbold{School. } A single school processes applications using an estimation policy $P$: for each student, upon observing information set $I$, the school forms an \textit{estimate} $\tilde q$ of their skill level $q$ via an estimation function $f_P(I)$. Information set $I$ always contains $\{\theta_k\}_{k=1}^{K-1}$ and $X$ (whether the student reports a score); if $X=1$, it further includes $\theta_K$. As in~\citet{garg2020standardized}, we study the \textit{Bayesian optimal} estimation policy $P_{BO}$, in which the student skill estimate is the posterior mean using all the information available:
$$\tilde q = f_{P_{BO}}(I) \triangleq \E[q | I].$$

We further consider non-Bayesian optimal estimation policies $P$, implemented by (potentially randomized) functions $f_P$.

What does the school observe for each student? In \Cref{sec:schooldoesnotknow}, we analyze the case in which the school does not observe whether a student has access to the test, $Z$; in \Cref{sec:schoolknows}, the school does know which students have access; i.e., $Z$ is in the information set $I$ used to estimate student skill. We find that this knowledge starkly changes the equilibrium and corresponding fairness guarantees.

What can the school require? We consider three requirement policies: no requirements (students are free to take and report the test unconstrained, $Z \geq Y \geq X$); reporting required conditional taking the test (if a student takes the test, then they must report their score, $Z \geq Y=X$);\footnote{Such a policy may be enforced with the help of testing agencies such as the College Board. \citet{Kannan_Niu_Roth_Vohra_2021} consider a similar policy.} or, in the case where the school knows $Z$, reporting required given access ($Z=Y=X$). We find that such requirements do not make a difference: the same results hold for all our settings, and so we do not include the requirement policy when denoting the school's policy $P$.

\parbold{Student decisions and equilibrium.} Pursuant to the requirement policy, students with access ($Z=1$) may decide to take the test ($Y=1$), with full knowledge of their skill $q$, noise distribution $\epsilon_K$, other features $\{\theta_k\}_{k=1}^{K-1}$ for ${k < K}$, and the school's estimation policy $P$. Students who take the test may, with knowledge of their test score $\theta_K$, decide to report it ($X=1$), in which case the school observes their  score $\theta_K$. Students make these decisions to maximize the school's estimate of their skill level. We denote the resulting decisions via the functions $Y(q, \{\theta_k\}_{k=1}^{K-1}, P)$ and $X(\{\theta_k\}_{k=1}^{K}, P)$.\footnote{It is simple to see that $X$ should not depend on $q$, given $\theta_K$. Once the test score is sampled from the student's skill level, the student's admission probability does not depend on $q$.}

We assume that the school knows the functions $X(\cdot),Y(\cdot)$.\footnote{For example, the school may observe values distributionally over time and learn them.} It can then use its knowledge of these functions to estimate the skill level of each student. For example, with Bayesian optimal estimation, if only students with $q \geq 1$ report their test score, then $\E[q | X = 1]\geq 1$. 

Thus, for students with access, we must calculate an equilibrium: student strategies to report a test score depend on the school's estimation, which in turn depends on the reporting decisions of students. An equilibrium is as follows.

\begin{definition}[Equilibrium] Let $\mathcal{T} = \{(Y,X) : Y \geq X, \text{(Y,X) obeys feature requirement policy}\}$ denote the action space of students with test access. Then, given estimation policy $P$ and a requirement policy, student decision functions $X(\cdot),Y(\cdot)$ and estimation function $f_p$ constitute an equilibrium if:

\begin{itemize}
    \item All decisions are feasible: $\left(Y(q, \{\theta_k\}_{k=1}^{K-1}, P), X(\{\theta_k\}_{k=1}^{K}, P)\right) \in \mathcal{T}$, for all $ \{\theta_k\}_{k=1}^{K-1}, q$. 
    \item Given a school's estimation function, each individual weakly prefers to follow their given action over other feasible actions: for all $ \{\theta_k\}_{k=1}^{K-1}, q$,
    \begin{align*}
        \left(Y(q, \{\theta_k\}_{k=1}^{K-1}, P), X(\{\theta_k\}_{k=1}^{K}, P)\right) \in {\arg\max}_{(y,x) \in \mathcal{T}} \E[\tilde q | Y=y, X=x] & 
    \end{align*}
    Where the expectation is performed over the (potentially random) school estimation function $f_P$, and, for $Y$, the randomness in the student's test score. 
    
    \item Estimation function $f_P$ follows its definition given the decision functions $X(\cdot),Y(\cdot)$. For example, suppose the school knows $Z$ and is using Bayesian optimal policy $P_{BO}$. Then, for students with access who do not report a score:
    \begin{align*}
        f_{P_{BO}}(\{\theta_k\}_{k=1}^{K-1}, X=0, Z = 1) &\triangleq \bbE_{X(\cdot)}[q |  \{\theta_k\}_{k=1}^{K-1}, X=0, Z = 1]\\
        &= \frac{\int_{\theta_K : X(\{\theta_k\}_{k=1}^{K}, P) = 0}\left( \bbE[q |  \{\theta_k\}_{k=1}^{K}]\Pr\left(\theta_K | \{\theta_k\}_{k=1}^{K-1}\right)\right) d\theta_K}
        {\Pr\left(\{\theta_K : X(\{\theta_k\}_{k=1}^{K}, P) = 0\} | \{\theta_k\}_{k=1}^{K-1}\right)}
    \end{align*}
\end{itemize}
    
\end{definition}

\subsection*{What does test-optional mean?}
The focus of this work is to understand how schools can meet their test-optional pledge. However, a prerequisite challenge is to formalize what ``test-optional'' even means, as it is left undefined by schools beyond the idea to  ``not penalize students for the absence of a standardized test score''~\cite{testoptional21}. Here, we define several potential notions, on the distribution of resulting skill estimates for students with and without test access. (Recall that, by assumption, neither true skill distribution $q$ nor feature distributions $\theta_k | q$ differ by test access). %

Our notions rely on ``distributional equality,'' denoted ($\eqdist$), because of the randomness in the process, both in the sampling of test scores given $q$ and in the estimation function $f_P$. Let $\tilde q | D,P$ denote the \textit{distribution} of $\tilde q$ given the information $D$ using policy $P$.

\begin{definition}[]
    \label{deffair1}
    Policy $P$ is \textbf{latent skill fair} if in every equilibrium, for all $q$,
    $$
    \tilde{q}|Z=1,\{\theta_k\}_{k=1}^{K-1},q,P \eqdist\tilde{q}|Z=0,\{\theta_k\}_{k=1}^{K-1}, q, P,
    $$
    i.e., if the distribution of skill estimates for students -- that share (latent) skill $q$ and (observable) other features $\{\theta_k\}_{k=1}^{K-1}$ -- with and without access are equal. 
\end{definition}

\begin{definition}[]
    \label{deffair2}
    Policy $P$ is \textbf{observably fair} if in every equilibrium, for all $\{\theta_k\}_{k=1}^{K-1}$,
    
    $$
    \tilde{q} | Z=1, \{\theta_k\}_{k=1}^{K-1},P \eqdist \tilde{q} | Z=0, \{\theta_k\}_{k=1}^{K-1}, P,
    $$
        i.e., if the distribution of skill estimates for students -- that share (observable) other features $\{\theta_k\}_{k=1}^{K-1}$ -- with and without access are equal.

\end{definition}

\begin{definition}[]
    \label{deffair3}
    Policy $P$ is \textbf{demographically fair} if in every equilibrium,
    $$
    \tilde{q}|Z=1,P\eqdist \tilde{q}|Z=0,P,
    $$
      i.e., if the distribution of skill estimates for students with and without access are equal.
\end{definition}

Each definition is less stringent than the previous, i.e., latent skill fairness $\implies$ observable fairness $\implies$ demographic fairness. The notions connect naturally to those common in the literature. Both latent and observable fairness are instances of \textit{individual} fairness, while demographic fairness is a \textit{group} fairness notion (however, one based on potentially \textit{unobserved} group membership). Observable fairness, in particular, aligns with the school's pledge to not penalize students without a test score: that two students with the same features $\theta_1$ to $\theta_{K-1}$ have the same distribution of scores.

A school can trivially achieve these notions if test scores are ignored altogether.

\begin{definition}[]
    Policy $P$ is \textbf{test-blank}, i.e., test scores play no role, if in every equilibrium,
    \begin{align*}
        \tilde q | \{\theta_k\}_{k=1}^{K-1}, P &\eqdist  \tilde q | \{\theta_k\}_{k=1}^{K-1},\theta_K, P &\forall \  \{\theta_k\}_{k=1}^{K}
    \end{align*}
\end{definition}
Of course, test-blank policies defeat the purpose of going test-optional. In this work, we thus ask whether it is possible to meet the fairness notions without being test-blank.

Note that these definitions all concern the estimates of the policy $P$ \textit{in equilibrium}. For example, there are two ways for a policy to be test-blank: either the estimation function $f_P$ ignores the test score (and whether the student reports it, $X$), or in equilibrium no student reports a test score. In either case, the definition is met, and test scores play no role in student estimates. Note also that for notational ease we suppress the requirement policy in the statement of each definition, as our results do not depend on which requirement policy is chosen.

\subsection*{Discussion}
Here, we briefly discuss some of our choices, with additional discussion in \Cref{sec:conclusion}.

First, note that our stated fairness definitions all condition on \textit{access} $Z$, as opposed to reporting $X$. However, all of our results, for each definition, also hold for the corresponding definition for which access $Z$ is replaced by reporting $X$; in fact, the corresponding results are easier to prove, as $X$ is always observed but $Z$ is potentially unobserved. In each proof, we explain the requisite changes needed for such a result. Our exposition centers $Z$ normatively, with the idea that schools should not reward those who strategically decide to not report.

Second, in this work we focus exclusively on how (and whether) a school can achieve the defined fairness notions. We do not explicitly consider other school goals, such as admitting students with the highest latent skill levels $q$ (``academic merit''). However, such a goal motivates our consideration of Bayesian estimation policy $P_{BO}$, which is optimal for academic merit when the students with the highest estimates are subsequently accepted; it also motivates us to find ways to meet the school's fairness goals without becoming test-blank, which similarly would worsen academic merit and be especially harmful for students without access to other informative signals with which to distinguish themselves~\cite{garg2020standardized}. 
Further note that our impossibility results hold for \textit{any} policy $P$ that might be chosen to maximize such other goals.

\section{When school does not know access status}
\label{sec:schooldoesnotknow}

In this section, we consider the case where the school does not observe (or does not use) whether a student has access to the test, $Z$. Instead, for each student, the school observes the first $K-1$ features, $\{\theta_k\}_{k=1}^{K-1}$, whether the student submits a score, $X$, and -- if the student submits the score ($X=1$) -- the score, $\theta_K$. This setting is consistent with the current policy adopted by many schools, which is that the school does not verify whether the student has access. %

Empirically with test-optional policies, some students who have taken the test choose not to report their score~\cite{morgan2016,robinsonmonks2005}; in particular, students with lower scores are more likely to withhold it. Our first result establishes that such withholding is the equilibrium result under Bayesian optimal estimation. As a result, the equilibrium does not satisfy any of our fairness definitions.

\begin{restatable}[Strategic withholding]{thm}{thmstrategicwithholding}
\label{thm:PBOnotknow}
    Suppose the school does not know access status and uses Bayesian optimal estimation $P_{BO}$ for all students. Then, some students with access do not report test scores. In particular, in every equilibrium:
    
    \begin{itemize}
        \item If the school does not require reporting conditional on taking the test, then $Y(\cdot) = 1$. For each $\thetaset$, there exists a finite test score threshold $\bar\theta_K$ such that an applicant reports their score, $X(\thetaset, \theta_K) = 1$, if and only if $\theta_K \geq \bar\theta_K$.
        
        \item If the school requires reporting conditional on taking the test ($Y=X$), then for each $\thetaset$, there exists a finite skill threshold $\bar q $ such that an applicant takes the test and reports their score, $Y(\thetaset, q) = 1$, if and only if $q \geq \bar q$.
    \end{itemize}
    
    Furthermore, $P_{BO}$ satisfies none of the fairness definitions.
    
\end{restatable}

To prove this theorem, we first argue that neither everyone report or no one report is equilibrium. We then show that if a fraction of the student reports their score (or takes the test), there should be a threshold for the test score (or the skill level) for reporting (or test-taking), and then we conclude by showing there exists such a threshold such that the equilibrium conditions hold. For the unfairness argument, we note that in this situation the estimated skill of a portion of students with access are guaranteed to strictly dominate those of their peers who do not have access, with others guaranteed to weakly dominate, so any fairness conditional on test access cannot be met.

The theorem establishes that the naive school strategy -- to optimally use whatever information is provided by the student -- induces students to strategically respond; as a result, the school satisfies none of our fairness criteria. More generally, the proof for unfairness is independent of the assumption that school does not know access status and the policy it uses. Instead, as long as the strategic effect exists, meeting fairness is impossible -- some of the students who have access are bound to benefit from their strategic advantage.

Furthermore, our next result establishes that latent and observable unfairness are unavoidable when the school does not know who has access, unless the school ignores the test score. This implies that being test-blank may be the only way to achieve a strategic-effect-free equilibrium in this situation, given the true information about students' access to tests is not known.

\begin{restatable}[Fairness impossibility]{thm}{thmnoaccessobsfair}
    Suppose the school does not know access status and uses an estimation policy $P$ that satisfies either latent skill or observable fairness. 
    Then, $P$ is test-blank.%

\end{restatable}
The proof proceeds via an unraveling argument. To satisfy those fairness notions and to avoid being test-blank, for any $X(\cdot)$ the policy $P$ must randomize skill estimates to equalize the estimate distributions. However, then, some students with $X(\thetaset, \theta_K)=1$ can strictly improve their estimates by no longer reporting their score. Reporting unravels until no student reports in equilibrium, which itself leads to the policy being test-blank. 

Our result does not establish whether demographic fairness is possible when the school does not know who has access -- that remains an open question. We conjecture that it cannot hold; however, the difficulty in the analysis is that policies can arbitrarily treat different other features, $\thetaset$, differently, such that demographic fairness may hold without the condition for observable fairness holding for any particular $\thetaset$. Note however that both of the above results (as all our results) also hold for fairness defined by conditioning on reporting $X$ -- even though the school observes it. 

This section establishes that achieving latent or observable fairness when the true access status is not known is generally impossible, if the school also wishes to make use of the test score. Both results in this section follow from the \textit{strategic} aspect of optional testing: students with low test scores can, by not reporting their score, masquerade as students without test access.

\section{When school knows access status}
\label{sec:schoolknows}

Given the general impossibility results in the previous section, we now turn to the setting in which the school knows whether a student has access to tests, $Z$.\footnote{While a school may never have true knowledge of $Z$ in practice, it may be able to approximate it by, for example, using submission information from others in the same socio-economic contexts. It may also ask students to credibly attest to a lack of access, or pursue other verification techniques. Our work provides an upper bound for how beneficial such knowledge may be.} 
Thus, the school now observes the first $K-1$ features, the access status, $Z$, and whether a students submits a score, $X$. If the student submits a score, $X=1$, then it further observes $\theta_K$; i.e., $I \triangleq (Z, \{\theta_k\}_{k=1}^{K-1}, X, \theta_K\mathbf{1}_{X})$. Note that this knowledge enables the school to follow different estimation policies for students with and without test access. We find that this knowledge substantially changes what is possible to achieve.

Our first result for this setting suggests why. In contrast to the setting without knowledge of $Z$, using a Bayesian optimal estimator of skill for students with access to tests results in all students with access submitting their score.

\begin{restatable}[Strategy-proofness]{lem}{lemstrategyaccess}
\label{lem:strategyaccess}
Suppose the school observes access $Z$ for each student and uses a policy $P$ that performs Bayesian optimal estimation on students with access; i.e., when $Z=1$,
$$f_P(I) = \bbE[q | I].$$

Then, the unique equilibrium is that all students with access take the test and report their scores:
\begin{align*}
Y(q, \{\theta_k\}_{k=1}^{K-1}, P) &= X(\{\theta_k\}_{k=1}^{K}, P) = 1.    & \forall q, \{\theta_k\}_{k=1}^{K}
\end{align*}
    
\end{restatable}

The proof proceeds as follows: the unraveling occurs in the opposite direction as in the case where access is not known. If some students with access do not report their score (or do not take the test), then for any fixed feature set $\{\theta_k\}_{k=1}^{K-1}$ there must exist a threshold $\bar\theta_K$ such that applicants do not submit a score if and only if $\theta_K < \bar\theta_K$ (or a true skill threshold $\bar q$ if students can strategize taking the test). However, under such threshold, there are always some students who have access and are currently not reporting (or taking the test) who can improve their estimated skill by reporting (or taking the test and then reporting). Thus any fraction of the students with access not reporting (or taking) is unstable, and so in equilibrium all students with access take the test. 

The result establishes that the school can induce all students with access to take the test and submit their scores, \textit{without having to mandate submission} -- simply by following a Bayesian optimal estimation policy. In other words, by simply following the policy that best makes use of given information, the school simultaneously induces students to report all available information. This result is the opposite of Theorem \ref{thm:PBOnotknow}, where the school did not know who had access and so such students could strategically withhold their scores.

What is different between the two settings? When the school does not know who has access, students with access but a low test score can disguise themselves as (potentially high true skill) students without test access. When the school does know, on the other hand, they cannot do so; if any fraction of students do not report their score, the school pools such students together and estimates the test scores for this pool -- and so there are always students who do not wish to pool with others who have also strategically with-held their scores.

Recall that in the proof of Theorem \ref{thm:PBOnotknow}, we established that if any students can strategically withhold their test score to improve their estimate, fairness is not possible. Lemma \ref{lem:strategyaccess} thus enables us to find fair policies by establishing that in the case where the school knows test access, Bayesian optimal estimation for applicants with access removes strategic incentives. Thus, for the rest of the section, we will assume that the school follows such a policy for applicants with test access, and focus on estimation for students without test access.

\subsection{Informational gaps cause unfairness}

Lemma \ref{lem:strategyaccess} eliminates \textit{strategic} concerns when the school uses Bayesian optimal estimation for those with test access. Thus, it may be tempting to conclude that it immediately follows that the fairness criteria hold. Our next result establishes the opposite: using Bayesian optimal estimation for those with test access  eliminates the possibility for latent skill fairness, regardless of what one does for those without access.

\begin{restatable}{prop}{propknowlatentfair}
\label{prop:knowlatentfair}
Suppose the school observes access $Z$. Consider any policy $P$ that performs Bayesian optimal estimation on students with access. Then, $P$ is not latent skill fair. %
\end{restatable}

The result follows from \textit{informational} differences: if the school uses the test score for those with access, then it cannot hope to equally accurately distinguish among students without access -- high true skill students without access have fewer opportunities to demonstrate their skill and receive a high skill estimate. In other words, if the school wishes to be latent skill fair, then it must artificially worsen its estimates for those with test access.\footnote{As an example of a policy that would be latent fair, suppose the school added additional noise to the estimates of students with access, such that the distributions equalize even conditioned on $q$.}

Thus, for the remainder of this section, we turn to studying observable and demographic fairness. However, even in this case, eliminating the strategic challenge does not guarantee that the school to satisfies these criteria; for example, our next result establishes that also using Bayesian optimal estimation for students without access does not satisfy these notions.

\begin{restatable}{prop}{propfairnessbasic}
\label{prop:knowobsfair}

    Suppose the school observes access status $Z$. Then, $P_{BO}$ -- the policy that uses Bayesian optimal estimation for all students -- is not observably or demographically fair.

\end{restatable}

The proof is straightforward and follows from Lemma \ref{lem:strategyaccess}: since every student with access reports their test score, Bayesian optimal estimation has more information on such students. One can then algebraically verify that the resulting distributions of $\tilde q$ differ.\footnote{In fact, the result follows immediately from our Lemma \ref{lem:strategyaccess} (which removes strategic concerns) and Lemma 3.1 in \citet{garg2020standardized}, which calculates the skill distributions using Bayesian-optimal estimation given a fixed feature set.}

\subsection{A non-Bayesian optimal but fair admissions process }
\label{sec:resampling}

Up to now, our results have been negative: even if the school knows who has access and uses this knowledge to eliminate strategic incentives, it does not meet the fairness criteria using existing policies. We now provide a positive result, a novel estimation policy that satisfies observable and demographic fairness. 

To overcome the issue identified in Proposition \ref{prop:knowobsfair}, a policy $P$ must randomize the estimated skill for students without access, in order to equalize the distributions of $\tilde{q}|Z=1,\{\theta_k\}_{k=1}^{K-1},P$ and $\tilde{q}|Z=0,\{\theta_k\}_{k=1}^{K-1},P$. With this randomization, students with the same observable features may receive different skill estimates. Consider the following policy.

\begin{definition} [Re-sampling policy]
    \label{def:pnb}
    The sampling estimation policy, $P_{S}$, is as follows.
    
    For students with access to tests, the school uses the Bayesian optimal method to estimate the student's true skill level; i.e., for $I\triangleq (Z=1,\{\theta_k\}_{k=1}^{K-1},X, \theta_K\mathbf{1}_{X})$:
    $$
f_{P_{S}}(I)=\mathbb{E}[q|I].
$$

    For students without access to tests, the school uses the first $K-1$ features $\{\theta_k\}_{k=1}^K$ to generate the distribution of the test score $\theta_K$, and then randomly draws a score for each student to add to the student's profile. Finally, the school estimates the student's true skill level based on the submitted features and the randomly drawn test score; i.e., for $I\triangleq (Z=0,\{\theta_k\}_{k=1}^{K-1})$:
$$
f_{P_{S}}(Z=0,\{\theta_k\}_{k=1}^{K-1})=\mathbb{E}[q|\{\theta_k\}_{k=1}^{K-1}, \tilde{\theta}_K]
$$
where
$$
\tilde{\theta}_K\sim \theta_K|\{\theta_k\}_{k=1}^{K-1}.
$$
\end{definition}

Then, we have the following.

\begin{restatable}{thm}{propfairnesssampling}
\label{thm:resamplingfair}

    Suppose the school observes access status $Z$. Then, the re-sampling policy $P_{S}$ is observably and demographically fair.

\end{restatable}

The proof follows from directly computing the induced skill estimate distributions. In addition to theoretically achieving our fairness notion, the policy can be described via a simple intuition. The school uses everything it knows about a student to generate the \textit{distribution} of what the student's skill estimate could have been had they had access to the test, and then it samples the student's estimate from the resulting distribution. 

This procedure is akin to Thompson sampling for bandits, in which an action is played according to the probability that it maximizes the reward. Here, if a school accepts all students with skill estimates above some threshold $\bar q$, then each student without access is accepted with probability equal to the probability they would receive a high enough score to be accepted had they taken the test; this probability is conditioned on everything the school knows about the student, $\{\theta_k\}_{k=1}^{K-1}$. %

Note that the re-sampling policy $P_S$ accepts some students who would not be accepted by the Bayesian optimal policy -- a school adopting $P_S$ would be recognizing that there are some students who do not have access to tests, but would excel if they did. And because it cannot identify the exact \textit{individual} students who would excel, it re-samples test scores to maintain distributional fairness for the \textit{group} (but still conditional on each individual's features $\thetaset$).

Together, the results in this section establish that a school can in theory overcome both the strategic and the informational gaps induced by test-optional policies. Overcoming strategic gaps requires using knowledge of who has access to the test (Lemma \ref{lem:strategyaccess}); overcoming informational gaps requires adopting a non-Bayesian optimal policy  (Theorem \ref{thm:resamplingfair}).

\section{Discussion}
\label{sec:conclusion}
Our model is stylized in several ways. One useful assumption is that the features are normally distributed with the true skill level as the mean. In practice, test scores are truncated and may be well fitted with truncated normal distributions,\footnote{See, e.g., SAT score distributions here: \url{https://collegereadiness.collegeboard.org/pdf/understanding-sat-scores.pdf}.%
} while scores of other features may exhibit far different distributions. This assumption is made for simplicity of exposition, following~\citet{garg2020standardized}. We expect our results to hold more generally outside the assumption; in our proofs, we do not use extensively the normal distribution itself, but rather properties such as symmetry and invariability to linear transforms. %

More consequential is our assumption that students seek to maximize their expected skill estimate $\tilde q$, as opposed to some (potentially non-linear) function of the estimate. For example, an alternative choice would be for students to maximize their probability of admissions, in a setting in which the school accepts the students with the highest skill estimates up to a capacity constraint. Doing so would not change our results (assuming appropriate tie-breaking for how students choose among equivalent strategies), except when the school mandates reporting of test scores given a student has taken it.  In that case, students with high scores in other features might -- even if the test increases their estimated skill in expectation -- not take the test to avoid the downside risk of a bad score resulting in a rejection. We believe that our choice is more natural, as in practice students may not know the admissions threshold but may know whether they are likely to do well on the test. 
Similarly, our notions and impossibility results are for strict (distributional) equality -- it is likely that policies differ in the \textit{magnitude} the fairness violations. We leave to future work to construct policies that, while maximizing some objective of interest, minimize the fairness violations.

Finally, we note that our focus on Bayesian optimal policies implicitly assumes that tests provide some useful information about students, and that schools process such information optimally (as opposed to, e.g., over-weighting what a low score says about a student). In this way, our framework provides an upper bound for when test-optional policies can be effective.

\parbold{Policy impact and future work} Following the empirical literature on test-optional policies, we analyze the fairness implications of such adoption. Our work shows that effective and fair implementation of such policies requires overcoming both \textit{strategic behavior} and \textit{informational gaps}. Both aspects are, to an extent, feasible to overcome: the school may learn about students' access status using submission information from others in the same socio-economic contexts, or by asking students to credibly attest to a lack of access; randomized admission may be achieved by drawing randomly from test scores observed in the group with test scores and similar features, or more informally by the school choosing to take ``risks'' on some students about which it has limited information. However, development of workable schemes in practice requires considerable work alongside practitioners. Our work provides guidance for the strategies that are likely to be effective.

More generally, our work is an example of a setting in which the strategic behavior of (some) agents introduces inequities that must be countered. Such effects are common in societal systems, though surprisingly rarely analyzed in the fair machine learning or mechanism design communities. We believe that our work provides an illustration of and lays the groundwork for such future work.

\bibliography{bib}

\appendix
\section{Proofs}

\thmstrategicwithholding*

\begin{proof}
\parbold{We begin with the claim under the assumption that reporting is not required conditional on taking the test.} We first restrict our analysis to a set of students with fixed features $\{\theta_k\}_{k=1}^{K-1}$. By a purely Bayesian estimation, the school estimates students' skill as follows: for students who report a score, the estimated skill is

$$
f_{P_{BO}}\left(\{\theta_{k}\}_{k=1}^{K},X=1\right)\triangleq \E [q|\{\theta_k\}_{k=1}^{K}]=\frac{\mu\sigma^{-2}+\sum_{k=1}^{K}\theta_k\sigma_k^{-2}}{\sigma^{-2}+\sum_{i=1}^{K}\sigma_k^{-2}},
$$
and for students who do not report a test score, the skill is estimated as follows:
\begin{align*}
    &f_{P_{BO}}\left(\{\theta_{k}\}_{k=1}^{K-1},X=0\right)\triangleq \E [q|\{\theta_k\}_{k=1}^{K-1}, X = 0]\\
    =&\E\left[q|X=0,\{\theta_k\}_{k=1}^{K-1}, Z=0\right]\text{Pr}\left(Z=0|X=0,\{\theta_{k}\}_{k=1}^{K-1}\right)\\
    &+\E\left[q|X=0,\{\theta_k\}_{k=1}^{K-1}, Z=1\right]\text{Pr}\left(Z=1|X=0,\{\theta_{k}\}_{k=1}^{K-1}\right)\\
    =&\frac{\mu\sigma^{-2}+\sum_{k=1}^{K-1}\theta_k\sigma_k^{-2}}{\sigma^{-2}+\sum_{i=1}^{K-1}\sigma_k^{-2}}\text{Pr}\left(Z=0|X=0,\{\theta_{k}\}_{k=1}^{K-1}\right)\\
    &+\frac{\int_{\theta_K : X(\{\theta_k\}_{k=1}^{K}, P_{BO}) = 0}\left( \bbE[q |  \{\theta_k\}_{k=1}^{K}]\Pr\left(\theta_K | \{\theta_k\}_{k=1}^{K-1}\right)\right) d\theta_K}
        {\Pr\left(\{\theta_K : X(\{\theta_k\}_{k=1}^{K}, P_{BO}) = 0\} | \{\theta_k\}_{k=1}^{K-1},Z=1\right)} \text{Pr}\left(Z=1|X=0,\{\theta_{k}\}_{k=1}^{K-1}\right)
\end{align*}
where the last equality is by definition of the equilibrium and properties of the normal distribution.

\parbold{Neither everyone reporting nor no one reporting is an equilibrium} We next show that neither ``everyone reports'' nor ``no one reports'' is a stable equilibrium for this set of students. Let us assume that everyone reports is the equilibrium, that is, $X(\{\theta_k\}_{k=1}^{K},P_{BO})=1, \ \forall \theta_K$. Then the latter quantity reduces to
$$
f_{P_{BO}}\left(\{\theta_{k}\}_{k=1}^{K-1},X=0\right)=\frac{\mu\sigma^{-2}+\sum_{k=1}^{K-1}\theta_k\sigma_k^{-2}}{\sigma^{-2}+\sum_{i=1}^{K-1}\sigma_k^{-2}}.
$$

However, there exist students with test scores
$$
\theta_K\le \sigma_K^2\left(\frac{(\sigma^{-2}+\sum_{k=1}^{K}\sigma_k^{-2})(\mu\sigma^{-2}+\sum_{k=1}^{K-1}\theta_k\sigma_k^{-2})}{\sigma^{-2}+\sum_{k=1}^{K-1}\sigma_k^{-2}}-\mu\sigma^{-2}-\sum_{k=1}^{K-1}\theta_k\sigma_k^{-2}\right),
$$
who would be better off if they chose not to report. So everyone reports is not stable.

Similarly, assume that no one reports is the equilibrium, we then have $X(\{\theta_k\}_{k=1}^{K},P_{BO})=0, \ \forall \theta_K$, and the integral would yield (the same result):
$$
f_{P_{BO}}\left(\{\theta_{k}\}_{k=1}^{K-1},X=0\right)=\frac{\mu\sigma^{-2}+\sum_{k=1}^{K-1}\theta_k\sigma_k^{-2}}{\sigma^{-2}+\sum_{i=1}^{K-1}\sigma_k^{-2}}.
$$

Again, there exist students with test scores 
$$
\theta_K\ge \sigma_K^2\left(\frac{(\sigma^{-2}+\sum_{k=1}^{K}\sigma_k^{-2})(\mu\sigma^{-2}+\sum_{k=1}^{K-1}\theta_k\sigma_k^{-2})}{\sigma^{-2}+\sum_{k=1}^{K-1}\sigma_k^{-2}}-\mu\sigma^{-2}-\sum_{k=1}^{K-1}\theta_k\sigma_k^{-2}\right),
$$
who would be better off if they chose to report. So no one reports is also not stable. Thus we know that if an equilibrium exists, it must be that a fraction of the students report.

\parbold{$X(\cdot)$ is non-decreasing in $\theta_K$} Following this, we argue that, if there exists such a situation that a fraction of the students report the score, then the function $X(\{\theta_k\}_{k=1}^{K}, P_{BO})$ is non-decreasing with respect to $\theta_K$. Let us assume that this does not hold, then there exists $\theta_K^1>\theta_K^2$ such that $0=X(\{\theta_k\}_{k=1}^{K-1}, \theta_K^1, P_{BO})< X(\{\theta_k\}_{k=1}^{K-1}, \theta_K^2, P_{BO})=1$. We then have
\begin{align*}
&f_{P_{BO}}\left(\{\theta_{k}\}_{k=1}^{K-1},X=0\right)\ge f_{P_{BO}}\left(\{\theta_{k}\}_{k=1}^{K-1},\theta_K^1,X=1\right)=\frac{\mu\sigma^{-2}+\sum_{k=1}^{K-1}\theta_k\sigma_k^{-2}+\theta_K^{1}\sigma_K^{-2}}{\sigma^{-2}+\sum_{i=1}^{K}\sigma_k^{-2}}\\
>&\frac{\mu\sigma^{-2}+\sum_{k=1}^{K-1}\theta_k\sigma_k^{-2}+\theta_K^{2}\sigma_K^{-2}}{\sigma^{-2}+\sum_{i=1}^{K}\sigma_k^{-2}}=f_{P_{BO}}\left(\{\theta_{k}\}_{k=1}^{K-1},\theta_K^2,X=1\right)\ge f_{P_{BO}}\left(\{\theta_{k}\}_{k=1}^{K-1},X=0\right)
\end{align*}
where all the equalities are by definition of Bayesian estimation, the first and last inequalities are by the strategy of the students, and the middle inequality is by assumption $\theta_K^1>\theta_K^2$. Here we observe a contradiction, thus we can assert that if there exists such a situation that a fraction of the students report the score, then the function $X(\{\theta_k\}_{k=1}^{K}, P_{BO})$ is non-decreasing with respect to $\theta_K$. Another way to say this is, there will be a test score threshold for reporting, $\bar{\theta}_K$, such that $X(\{\theta_k\}_{k=1}^{K}, P_{BO})=1$ for $\theta_K\ge \bar{\theta}_K$ and $X(\{\theta_k\}_{k=1}^{K}, P_{BO})=0$ otherwise. 

\parbold{There exists a equilibrium threshold} which defines the function $X$ for this set of students. To show this, we first rewrite $f_{P_{BO}}\left(\{\theta_{k}\}_{k=1}^{K-1},X=0\right)$ in terms of $\bar{\theta}_K$. Let us assume, without loss of generality, that within this set of students with fixed $\{\theta_k\}_{k=1}^{K}$, the fraction of students who have access to tests is $C\in (0,1)$. Then we have that of this fraction of students with access, a further fraction of 
$$
\int_{\theta_K}X(\{\theta_k\}_{k=1}^{K-1},\theta_K, P_{BO})\text{Pr}\left(\theta_K|\{\theta_k\}_{k=1}^{K-1}\right)d\theta_K=\int_{\bar{\theta}_K}^{\infty}\text{Pr}(\theta_K|\{\theta_k\}_{k=1}^{K-1})d\theta_K=\text{Pr}[\theta_K\ge \bar{\theta}_K|\{\theta_k\}_{k=1}^{K-1}]\triangleq A(\bar{\theta}_K)
$$
will report the score. The first equality is because of the threshold of $X$, and the second is by property of the cumulative distribution. Note that $A(\bar{\theta}_K)$ is continuous in $\bar{\theta}_K$. With this, we further have:
$$
\text{Pr}\left(Z=1|X=0,\{\theta_{k}\}_{k=1}^{K-1}\right)=\frac{CA(\bar{\theta}_K)}{CA(\bar{\theta}_K)+1-C}
$$
and
$$
\text{Pr}\left(Z=0|X=0,\{\theta_{k}\}_{k=1}^{K-1}\right)=\frac{1-C}{CA(\bar{\theta}_K)+1-C}
$$

Next we have as a direct result
$$
\Pr\left(\{\theta_K : X(\{\theta_k\}_{k=1}^{K}, P_{BO}) = 0\} | \{\theta_k\}_{k=1}^{K-1},Z=1\right)=\Pr\left(\{\theta_K<\bar{\theta}_K\} | \{\theta_k\}_{k=1}^{K-1}\right)=1-A(\bar{\theta}_K)
$$

For the integral part, we leave it largely unchanged:
$$
\int_{\theta_K : X(\{\theta_k\}_{k=1}^{K}, P_{BO}) = 0}\left( \bbE[q |  \{\theta_k\}_{k=1}^{K}]\Pr\left(\theta_K | \{\theta_k\}_{k=1}^{K-1}\right)\right) d\theta_K=\int_{-\infty}^{\bar{\theta}_K}\left( \bbE[q |  \{\theta_k\}_{k=1}^{K}]\Pr\left(\theta_K | \{\theta_k\}_{k=1}^{K-1}\right)\right) d\theta_K
$$

Assembling them together, we have
\begin{align*}
    &f_{P_{BO}}\left(\{\theta_{k}\}_{k=1}^{K-1},X=0\right)\\
    =&\frac{\mu\sigma^{-2}+\sum_{k=1}^{K-1}\theta_k\sigma_k^{-2}}{\sigma^{-2}+\sum_{i=1}^{K-1}\sigma_k^{-2}} \frac{1-C}{CA(\bar{\theta}_K)+1-C}\\
    &+\frac{\int_{-\infty}^{\bar{\theta}_K}\left( \bbE[q |  \{\theta_k\}_{k=1}^{K}]\Pr\left(\theta_K | \{\theta_k\}_{k=1}^{K-1}\right)\right) d\theta_K}{1-A(\bar{\theta}_K)}\times \frac{CA(\bar{\theta}_K)}{CA(\bar{\theta}_K)+1-C}
\end{align*}

An important observation is that, for any finite $\bar{\theta}_K$, that is $A(\bar{\theta}_K)\in(0,1)$, the above quantity as a function of $\bar{\theta}_K$ is continuous. The continuity of the integral part is justified because the integrand is always finite. 

Based on previous discussion, we know that when $\bar{\theta}_K\rightarrow \infty$, that is no one reports, 
$$
f_{P_{BO}}\left(\{\theta_{k}\}_{k=1}^{K-1},\bar{\theta}_K,X=1\right)>f_{P_{BO}}\left(\{\theta_{k}\}_{k=1}^{K-1},X=0\right),
$$
and when $\bar{\theta}_K\rightarrow -\infty$, that is when everyone reports,
$$
f_{P_{BO}}\left(\{\theta_{k}\}_{k=1}^{K-1},\bar{\theta}_K,X=1\right)<f_{P_{BO}}\left(\{\theta_{k}\}_{k=1}^{K-1},X=0\right).
$$

Since $f_{P_{BO}}\left(\{\theta_{k}\}_{k=1}^{K-1},\bar{\theta}_K,X=1\right)$ is linear in $\bar{\theta}_K$, and $f_{P_{BO}}\left(\{\theta_{k}\}_{k=1}^{K-1},X=0\right)$ is continuous in $\bar{\theta}_K$, we conclude that there must exist at least one finite $\bar{\theta}_K^*$, such that 
$$
f_{P_{BO}}\left(\{\theta_{k}\}_{k=1}^{K-1},\bar{\theta}_K^*,X=1\right)=f_{P_{BO}}\left(\{\theta_{k}\}_{k=1}^{K-1},X=0\right).
$$

Finally, note that the resulting function with $\bar{\theta}_K^*$ satisfies the equilibrium requirements; For anyone with test score $\theta_K<\bar{\theta}_K^*$, they are currently not reporting, and 
$$
f_{P_{BO}}\left(\{\theta_{k}\}_{k=1}^{K-1},\theta_K,X=1\right)<f_{P_{BO}}\left(\{\theta_{k}\}_{k=1}^{K-1},X=0\right),
$$
so they would not like to change their behavior to reporting; and vice versa for everyone with test score $\theta_K\ge \bar{\theta}_K^*$.

Extending the argument to all students with all possible sets of features $\{\theta_{k}\}_{k=1}^{K-1}$, we get the desired result: there exists an equilibrium in which for any set of students with features $\{\theta_{k}\}_{k=1}^{K-1}$, there exists a threshold for test score $\bar{\theta}_K^*$, such that anyone below this test score does not submit the score, and anyone at or above this threshold submits the score.

\parbold{Next, it is simple to carry the above proof to the case where reporting is required conditional on taking the test. }We briefly sketch the proof as follows. Note that the estimation function $f_{P_{BO}}$ should be rewritten in terms of $Y(\cdot)$ and $q$. For students who takes the test, the estimated skill is
$$
f_{P_{BO}}\left(\{\theta_{k}\}_{k=1}^{K-1},Y(\thetaset,q,P_{BO})=X=1\right)\triangleq \E [q|\{\theta_k\}_{k=1}^{K-1},\theta_K|q]=\frac{\mu\sigma^{-2}+\sum_{k=1}^{K-1}\theta_k\sigma_k^{-2}+\sigma_K^{-2}\left(\theta_K|q\right)}{\sigma^{-2}+\sum_{i=1}^{K}\sigma_k^{-2}},
$$
where we abused the notations a bit: here $q$ is not observed by the school, and $\theta_K$ is not known and random to the students at the time of his decision to take the test. Since this skill estimate is random at the time of decision, what the student would really care about is actually the expectation of the above amount, which is 
$$
\E\left[f_{P_{BO}}\left(\{\theta_{k}\}_{k=1}^{K-1},Y(\thetaset,q,P_{BO})=X=1\right)\right]=\frac{\mu\sigma^{-2}+\sum_{k=1}^{K-1}\theta_k\sigma_k^{-2}+q\sigma_K^{-2}}{\sigma^{-2}+\sum_{i=1}^{K}\sigma_k^{-2}}
$$

For students who do not report a test score, the skill is estimated as follows:
\begin{align*}
    &f_{P_{BO}}\left(\{\theta_{k}\}_{k=1}^{K-1},Y(\thetaset,q,P_{BO})=X=0\right) \triangleq \E [q|\{\theta_k\}_{k=1}^{K-1}, Y=X = 0]\\
    =&\E\left[q|Y=X=0,\{\theta_k\}_{k=1}^{K-1}, Z=0\right]\text{Pr}\left(Z=0|Y=X=0,\{\theta_{k}\}_{k=1}^{K-1}\right)\\
    &+\E\left[q|Y=X=0,\{\theta_k\}_{k=1}^{K-1}, Z=1\right]\text{Pr}\left(Z=1|Y=X=0,\{\theta_{k}\}_{k=1}^{K-1}\right)\\
    =&\frac{\mu\sigma^{-2}+\sum_{k=1}^{K-1}\theta_k\sigma_k^{-2}}{\sigma^{-2}+\sum_{i=1}^{K-1}\sigma_k^{-2}}\text{Pr}\left(Z=0|Y=X=0,\{\theta_{k}\}_{k=1}^{K-1}\right)\\
    &+\frac{\int_{q : Y(\{\theta_k\}_{k=1}^{K-1},q, P_{BO}) = 0}\int_{\theta_K}\left( \bbE[q |  \{\theta_k\}_{k=1}^{K}]\Pr(\theta_K|q)\Pr\left(q | \{\theta_k\}_{k=1}^{K-1}\right)\right) d\theta_Kdq}
        {\Pr\left(\{q : Y(\{\theta_k\}_{k=1}^{K-1},q, P_{BO}) = 0\} | \{\theta_k\}_{k=1}^{K-1},Z=1\right)}\\
        &\ \ \times \text{Pr}\left(Z=1|Y=X=0,\{\theta_{k}\}_{k=1}^{K-1}\right)
\end{align*}
where the last equality is by definition of the equilibrium and properties of the normal distribution.

We can then show that neither everyone takes the test and reports nor no one takes the test and reports are stable, so if an equilibrium exists, it must be that a fraction of students report. Under such situation, the function $Y(\thetaset,q,P_{BO})$ would be increasing with respect to $q$, so that there is a threshold of $q$ for test-taking. We can then finish by showing that there exists such a threshold which defines the function $Y(\cdot)$ for each $\thetaset$, that satisfies equilibrium. The technique is the same, by noting that $f_{P_{BO}}\left(\{\theta_{k}\}_{k=1}^{K-1},Y(\thetaset,q,P_{BO})=X=0\right)$ is continuous in $q$ and $\E\left[f_{P_{BO}}\left(\{\theta_{k}\}_{k=1}^{K-1},Y(\thetaset,q,P_{BO})=X=1\right)\right]$ is linear in  $q$, and that their relation does not agree when $q$ goes to plus or minus infinity.

\parbold{Finally, to conclude that it is unfair according to all definitions,} we note that in this regime, every student who has access to test can act strategically, so that he has an estimated skill that weakly dominates another student who is completely the same otherwise, but does not have access to test. Furthermore, there always exist some students who have access that strictly dominate their counterparts, otherwise the situation where no one report the score (or take the test) should be stable. Thus,  $\tilde q | Z=1,\thetaset$ strictly stochastically dominates $\tilde q | Z=0,\thetaset$ for every $\thetaset$. This means any equality in distribution that is conditional on $Z$ cannot hold, hence all three fairness criteria fail. As a side remark, if we were considering fairness conditioned on $X$, the same argument still follows: all students with $X=1$ weakly benefit from this decision (and some strictly benefit), while a portion of the students with $X=0$, namely those without access, does not benefit at all.

\end{proof}

\thmnoaccessobsfair*

\begin{proof}
    \parbold{We begin the proof for the observably fair part.} First we consider the case when reporting is not required conditional on taking the test, and students choose whether to report the score after taking the test. In this case, to meet observable fairness:
    $$
    \tilde{q} | Z=0, \{\theta_k\}_{k=1}^{K-1},P \eqdist \tilde{q} | Z=1, \{\theta_k\}_{k=1}^{K-1}, P, \quad \forall\  \{\theta_k\}_{k=1}^{K-1},
    $$
    there are two approaches. 
    
    The first approach is by eliminating the dependence on $\theta_K$ on the right hand side, which is choosing a policy $P$ to ignore the test score completely in the estimation process (which is test blank):
    $$
        f_P\left(\{\theta_k\}_{k=1}^{K-1}\right)\eqdist  f_P\left(\{\theta_k\}_{k=1}^{K-1},\theta_K\right),\quad \forall \  \{\theta_k\}_{k=1}^{K}
    $$
    then trivially, observable fairness is met by definition of $\tilde{q}$. Note that if $P$ is test blank, then no students who currently report their score have the incentive to stop reporting, since their estimated skill is independent of their test score, so if they stop reporting, it remains the same. This holds for any student who currently does not report as well.
    
    The second approach is by adding a replacement of $\theta_K$ on the left hand side to try to equalize the distributions, which can be achieved by introducing test score sampling similar to that discussed in section \ref{sec:resampling}. However, since the school does not know exactly who has access, it can only build on who reports a test score. For a fixed set of $\{\theta_k\}_{k=1}^{K-1}$, let us denote:
    \begin{align*}
    &\tilde{q}|X=1,\{\theta_k\}_{k=1}^{K-1}, P \triangleq D_{P,1},\\
    & \tilde{q}|X=0,\{\theta_k\}_{k=1}^{K-1}, P \triangleq D_{P,0}.
    \end{align*}
    Then we claim that for any fraction $\lambda>0$ of students who report their score, any policy $P$ that tries to equalize the distributions will make the situation unstable. For any $\lambda\in (0,1]$, we have:
    \begin{align*}
    &\tilde{q} | Z=0, \{\theta_k\}_{k=1}^{K-1},P \eqdist D_{P,0}\\
    \eqdist\ &\tilde{q} | Z=1, \{\theta_k\}_{k=1}^{K-1}, P \eqdist \lambda D_{P,1}+(1-\lambda)D_{P,0}  \\
    \Rightarrow &D_{P,1}=D_{P,0}.
    \end{align*}
    This implies that $\hat{\theta}_K|X=0\eqdist \theta_K|X=1$, which means that in the resampling procedure used by $P$, $\hat{\theta}_K$ is sampled from a distribution $\mathcal{D}_{K}$ that is exactly the observed distribution of $\theta_K$ with those who submit the score. Now for any student who currently submits the score but has a score $\theta_K$ that satisfies $\theta_K\le \bar{\theta}_K$, where $\bar{\theta}_K$ is the mean of $\mathcal{D}_K$ he would want to stop reporting since
    \begin{align*}
        \tilde{q}|X=1,\{\theta_k\}_{k=1}^{K-1},\theta_K&=\frac{\mu\sigma^{-2}+\sum_{k=1}^{K}\theta_k\sigma_k^{-2}}{\sigma^{-2}+\sum_{i=1}^{K}\sigma_k^{-2}}\\
        &\le \frac{\mu\sigma^{-2}+\sum_{k=1}^{K-1}\theta_k\sigma_k^{-2}+\bar{\theta}_K\sigma_K^{-2}}{\sigma^{-2}+\sum_{i=1}^{K}\sigma_k^{-2}}=\E[\tilde{q}|X=0,\{\theta_k\}_{k=1}^{K-1}],
    \end{align*}
    showing that such a situation is unstable. Note that such a student always exists, following the definition of the mean of the distribution $\mathcal{D}_K$.

    Extending the above argument to all set of features $\{\theta_k\}_{k=1}^{K-1}$, we can conclude that in this case, any policy that tries to be observably fair can only ignore the test score in the estimation process, or lead to no one reporting in equilibrium, hence is test-blank.

    \vspace{.2cm}

    Next we consider the second case when reporting is required conditional on taking the test, and students choose whether to take the test. In this case, there are equally two approaches.

    The first approach is by using a policy $P$ that ignores the test score in estimation. Following the definition, a test-blank policy $P$ would also guarantee observable fairness, regardless of anyone reporting.

    The second approach is by introducing resampling techniques. We note that an analogue of the above proof technique still holds. For a fixed set of $\{\theta_k\}_{k=1}^{K-1}$, let us still denote:
    \begin{align*}
        &\tilde{q}|X=1,\{\theta_k\}_{k=1}^{K-1}, P \triangleq D_{P,1},\\
        & \tilde{q}|X=0,\{\theta_k\}_{k=1}^{K-1}, P \triangleq D_{P,0}.
    \end{align*}
    Then we claim that for any fraction $\lambda>0$ of students who report their score, any policy $P$ that tries to equalize the distributions will make the situation unstable. For any $\lambda\in (0,1]$, we have:
    \begin{align*}
    &\tilde{q} | Z=0, \{\theta_k\}_{k=1}^{K-1},P \eqdist D_{P,0}\\
    \eqdist\ &\tilde{q} | Z=1, \{\theta_k\}_{k=1}^{K-1}, P \eqdist \lambda D_{P,1}+(1-\lambda)D_{P,0}  \\
    \Rightarrow &D_{P,1}=D_{P,0}.
    \end{align*}
    This implies that $\hat{\theta}_K|X=0\eqdist \theta_K|X=1$, which means that in the resampling procedure used by $P$, $\hat{\theta}_K$ is sampled from a distribution $\mathcal{D}_{K}$ that is exactly the observed distribution of $\theta_K$ with those who submit the score. Now for any student who currently chooses to take the test, but has a latent skill level $q$ that satisfies $q\le \bar{\theta}_K$, where $\bar{\theta}_K$ is the mean of $\mathcal{D}_K$, we get that:
    \begin{align*}
        \E[\tilde{q}|Y=X=1,\{\theta_k\}_{k=1}^{K-1},q]&=\frac{\mu\sigma^{-2}+\sum_{k=1}^{K-1}\theta_k\sigma_k^{-2}+q\sigma_K^{-2}}{\sigma^{-2}+\sum_{i=1}^{K}\sigma_k^{-2}}\\
        &\le \frac{\mu\sigma^{-2}+\sum_{k=1}^{K-1}\theta_k\sigma_k^{-2}+\bar{\theta}_K\sigma_K^{-2}}{\sigma^{-2}+\sum_{i=1}^{K}\sigma_k^{-2}}=\E[\tilde{q}|Y=X=0,\{\theta_k\}_{k=1}^{K-1}],
    \end{align*}
    so he would like to stop taking the test, showing that such a situation is not stable. Note that such a student always exists, this is because the distribution of $q$ in the cohort of students who take the test has the same mean as $\mathcal{D}_K$, the observed distribution of $\theta_K$. So in expectation, all students in this cohort who has a skill level below average should want to stop taking the test.

    Again, extending the above argument to all set of features $\{\theta_k\}_{k=1}^{K-1}$, we can conclude that in this second case, any policy that tries to be observably fair can only ignore the test score in estimation or lead to no one reporting in equilibrium, hence is test-blank.
    
    \parbold{Recalling that latent skill fairness is stricter than observable fairness, the statement for latent skill fair follows immediately.}

    To migrate the above proof to the case where reporting is required conditional on testing, simply replace the integral over $\theta_K$ by a double integral over $\theta_K|q$ and $q$, the other arguments are the same, and the result is that the unique equilibrium is no one takes the test and reports the score.
    
    To sum up, any policy that satisfies demographic fairness will be either test-blank, or lead to no one reporting in equilibrium.
    
    As a side remark, all the proof relies on equalizing the distribution conditional on $X$, thus the proof also works for fairness conditional on $X$.
\end{proof}

\begin{customlem}{4.1}[Strategy-proofness]
Suppose the school observes access $Z$ for each student and uses a policy $P$ that performs Bayesian optimal estimation on students with access; i.e., when $Z=1$,
$$f_P(I) = \bbE[q | I].$$

Then, the unique equilibrium is that all students with access take the test and report their scores:
\begin{align*}
Y(q, \{\theta_k\}_{k=1}^{K-1}, P) &= X(\{\theta_k\}_{k=1}^{K}, P) = 1.    & \forall q, \{\theta_k\}_{k=1}^{K}
\end{align*}
    
\end{customlem}

\begin{proof}
    \parbold{First case: when reporting is not required after taking the test, and students choose whether to report the score after taking the test.} In this setting, we assume that all students with test access take the test, i.e. $Y(\cdot)=1$. Thus, $\theta_K$ is available to all students who have access, and they now have to decide whether to report it or not, with the knowledge of all their features, the estimation policy, and the decisions and scores of all other students (as assumed in the model section).

    \parbold{Monotonicity of $X(\cdot)$} First, for a cohort of students with a fixed set of features $\{\theta_k\}_{k=1}^{K-1}$, we claim that the function $X\left(\{\theta_k\}_{k=1}^{K-1},\theta_K, P\right)$ is monotonically non-decreasing in $\theta_K$: $X\left(\{\theta_k\}_{k=1}^{K-1}, \theta_K, P\right) \geq X\left(\{\theta_k\}_{k=1}^{K-1}, \theta_K', P\right)$ if $\theta_K > \theta_K'$. 
    
    We prove this by contradiction. Suppose there exists $\theta_K^1 > \theta_K^2$ such that $X\left(\{\theta_k\}_{k=1}^{K-1}, \theta_K^1, P\right) < X\left(\{\theta_k\}_{k=1}^{K-1}, \theta_K^2, P\right)$. By the assumption that a student will choose to report test score if and only if not reporting hurts the estimated skill level, that is: 
    \begin{align*}
        &X\left(\{\theta_k\}_{k=1}^{K-1}, \theta_K, P\right)=1 \Leftrightarrow \E\left[q\vert\{\theta_k\}_{k=1}^{K-1},X=1,\theta_K\right]\ge \E\left[q\vert\{\theta_k\}_{k=1}^{K-1},X=0\right],\\
        &X\left(\{\theta_k\}_{k=1}^{K-1}, \theta_K, P\right)=0 \Leftrightarrow \E\left[q\vert\{\theta_k\}_{k=1}^{K-1},X=1,\theta_K\right]< \E\left[q\vert\{\theta_k\}_{k=1}^{K-1},X=0\right].
    \end{align*}
    Since $X\left(\{\theta_k\}_{k=1}^{K-1}, \theta_K^1, P\right) < X\left(\{\theta_k\}_{k=1}^{K-1}, \theta_K^2, P\right)$ and $X(\cdot)$ has range $\{0,1\}$ we have $X\left(\{\theta_k\}_{k=1}^{K-1}, \theta_K^1, P\right)=0$ and $X\left(\{\theta_k\}_{k=1}^{K-1}, \theta_K^2, P\right)=1$, this means that:
    \begin{align*}
        \E\left[q\vert \{\theta_k\}_{k=1}^{K-1}, X=1, \theta_K^1 \right] &< \E\left[q\vert\{\theta_k\}_{k=1}^{K-1},X=0\right]\\
        &\le \E\left[q\vert \{\theta_k\}_{k=1}^{K-1}, X=1, \theta_K^2 \right].
    \end{align*}
    However, under Bayesian optimal estimation, for fixed $\{\theta_k\}_{k=1}^{K-1}$ and $\theta_K^1>\theta_K^2$, we have 
    \begin{align*}
    \E\left[q\vert \{\theta_k\}_{k=1}^{K-1}, X=1, \theta_K^1 \right]
    &=\frac{\mu\sigma^{-2}+\sum_{k=1}^{K-1}\theta_k\sigma_k^{-2}+\theta_K^1\sigma_K^{-2}}{\sigma^{-2}+\sum_{i=1}^{K}\sigma_k^{-2}}\\
    &> \frac{\mu\sigma^{-2}+\sum_{k=1}^{K-1}\theta_k\sigma_k^{-2}+\theta_K^2\sigma_K^{-2}}{\sigma^{-2}+\sum_{i=1}^{K}\sigma_k^{-2}}
    = \E\left[q\vert \{\theta_k\}_{k=1}^{K-1}, X=1, \theta_K^2 \right]
    \end{align*}
    thus it is a contradiction. So we establish the monotonicity of $X(\cdot)$ given $\{\theta_k\}_{k=1}^{K-1}$.

    \parbold{Threshold induces students to change behavior} Next, if not all students report their test score, monotonicity of $X(\cdot)$ given $\{\theta_k\}_{k=1}^{K-1}$ implies that there exists a threshold $\bar{\theta}_K$ such that if $\theta_K\ge \bar{\theta}_K$, $X\left(\{\theta_k\}_{k=1}^{K-1}, \theta_K, P\right)=1$ and if $\theta_K<\bar{\theta}_K$, $X\left(\{\theta_k\}_{k=1}^{K-1}, \theta_K, P\right)=0$. The threshold is observable to the school, thus for those students who do not report their test score, the school now knows that their test score is below the threshold, so a Bayesian optimal estimation policy will estimate their skill level to be:
    $$
    \E\left[q|\{\theta_k\}_{k=1}^{K-1},X=0\right]=\frac{\int_{-\infty}^{\bar{\theta}_K}\E\left[q\vert \{\theta_k\}_{k=1}^{K-1}, X=1, \theta_K=\gamma \right]\text{Pr}\left(\theta_K=\gamma| \{\theta_k\}_{k=1}^{K-1}\right)d\gamma}{\int_{-\infty}^{\bar{\theta}_K}\text{Pr}\left(\theta_K=\gamma|\{\theta_k\}_{k=1}^{K-1}\right)d\gamma},
    $$
    where in this case $\theta_K|\{\theta_k\}_{k=1}^{K-1}\sim \mathcal{N}\left(\frac{\mu\sigma^{-2}+\sum_{k=1}^{K-1}\theta_k\sigma_k^{-2}}{\sigma^{-2}+\sum_{i=1}^{K-1}\sigma_k^{-2}},\sigma_K^2+\frac{1}{\sigma^{-2}+\sum_{i=1}^{K-1}\sigma_k^{-2}}\right)$. By strict monotonicity of the expectation $\E\left[q\vert \{\theta_k\}_{k=1}^{K-1}, X=1, \theta_K \right]=\frac{\mu\sigma^{-2}+\sum_{k=1}^{K}\theta_k\sigma_k^{-2}}{\sigma^{-2}+\sum_{i=1}^{K}\sigma_k^{-2}}$ with respect to $\theta_K$, we get that there exists $\tilde{\theta}_{K}<\bar{\theta}_K$ such that
    $$
    \E\left[q|\{\theta_k\}_{k=1}^{K-1},X=1, \tilde{\theta}_K\right]= \E\left[q|\{\theta_k\}_{k=1}^{K-1},X=0\right].
    $$
    
    Then we observe that for students with test score $\tilde{\theta}_{K}\le\theta_K<\bar{\theta}_K$, they are currently not reporting their scores, but would like to report since $\E\left[q|\{\theta_k\}_{k=1}^{K-1},X=1, \theta_K\right]\ge \E\left[q|\{\theta_k\}_{k=1}^{K-1},X=1, \tilde{\theta}_K\right]=\E\left[q|\{\theta_k\}_{k=1}^{K-1},X=0\right]$. This shows that such a situation is not stable.

    Notice that the above argument holds whenever not all students report their score, since then we are bound to have a threshold of $\bar{\theta}_K$ for reporting, and the rest of the argument follows. Thus, for a group of students with fixed $\{\theta_k\}_{k=1}^{K-1}$, the unique equilibrium is that all of them report their test score. Further notice that the argument can be extended to all $\{\theta_k\}_{k=1}^{K-1}$, so the unique equilibrium for all students, regardless of their other features, is that all of them report their test score. This concludes the proof for the case where reporting is optional conditioning on taking the test.

    \vspace{.2cm}

    \parbold{Second case: when reporting is required after taking the test, and students choose whether to take the test.} In this setting, we similarly begin the proof by restricting our argument to a cohort of students with a fixed set of features $\{\theta_k\}_{k=1}^{K-1}$, and claim that $Y(\{\theta_k\}_{k=1}^{K-1}, q, P) \geq Y(\{\theta_k\}_{k=1}^{K-1}, q', P)$ if $q > q'$, i.e. $Y(\{\theta_k\}_{k=1}^{K-1}, q, P)$ is monotonically non-decreasing in $q$ given $\{\theta_k\}_{k=1}^{K-1}$.

    \parbold{Monotonicity of $Y(\cdot)$} We also prove this by contradiction. Suppose there exists $q_1>q_2$ such that $0=Y(\{\theta_k\}_{k=1}^{K-1}, q_1, P) < Y(\{\theta_k\}_{k=1}^{K-1}, q_2, P)=1$. Recall that we defined the strategy of students to be maximize the estimated skill level. Here we further interpret this to be that a student will choose to take the test if this has a higher probability of leading to a better result than a worse one. Let us assume that there exists $\theta'_K$ such that
    \begin{equation}
        \label{eq:behavior_must_report}
        \E\left[q|\{\theta_k\}_{k=1}^{K-1},\theta'_K, Y=X=1\right]= \E\left[q|\{\theta_k\}_{k=1}^{K-1},Y=0\right],
    \end{equation}
    then obtaining any test score higher than $\theta'_K$ will improve the estimated skill level by monotonicity of the expectation. Thus, we can characterize the behavior of the students as:
    \begin{align*}
        Y(\{\theta_k\}_{k=1}^{K-1}, q, P)=1 \Leftrightarrow  \E(\theta_K|q)\ge \theta_K',\\
        Y(\{\theta_k\}_{k=1}^{K-1}, q, P)=0 \Leftrightarrow  \E(\theta_K|q)< \theta_K'.
    \end{align*}

    By assumption, we then have:
    $$
    \E(\theta_K|q_1)<\theta_K'\le \E(\theta_K|q_2).
    $$
    However, since $\E[\theta_K|q]=q$, we have $\E(\theta_K|q_1)=q_1>q_2=\E(\theta_K|q_2)$ and there is a contradiction. So we get that $Y(q, \{\theta_k\}_{k=1}^{K-1}, P)$ is indeed non-decreasing in $q$ given $\{\theta_k\}_{k=1}^{K-1}$.

    \parbold{Threshold induces students to change behavior} Now if not all students take the test, then by monotonicity of $Y(\cdot)$, we again know that there would be a threshold $\bar{q}$ such that if $q>\bar{q}$, $Y(\{\theta_k\}_{k=1}^{K-1}, q, P)=1$ and if $q<\bar{q}$, $Y(\{\theta_k\}_{k=1}^{K-1}, q, P)=0$. We notice that, with this threshold, the observed test score would follow the distribution of $\theta_K|\{\theta_k\}_{k=1}^{K-1}, q>\bar{q}$, which is:
    \begin{align*}
        \text{Pr}\left(\theta_K=\gamma|\{\theta_k\}_{k=1}^{K-1}, q>\bar{q}\right)=\frac{\int_{\bar{q}}^{\infty}\text{Pr}\left(\theta_K=\gamma|\theta_K\sim \mathcal{N}(\pi,\sigma_K^2)\right)\text{Pr}\left(q=\pi| q\sim \mathcal{N}\left(\hat{\mu}, \hat{\sigma}^2\right) \right)d\pi}{\int_{\bar{q}}^{\infty}\text{Pr}\left(q=\pi| q\sim \mathcal{N}\left(\hat{\mu}, \hat{\sigma}^2\right) \right)d\pi}.
    \end{align*}
    where $\hat{\mu}=\frac{\mu\sigma^{-2}+\sum_{k=1}^{K-1}\theta_k\sigma_k^{-2}}{\sigma^{-2}+\sum_{i=1}^{K-1}\sigma_k^{-2}}$ and $\hat{\sigma}^2=\frac{1}{\sigma^{-2}+\sum_{i=1}^{K-1}\sigma_{k}^{-2}}$.

    Note that we assume that the parameters $\sigma^2$ and $\sigma_1^2,...,\sigma_K^2$ are known, and so the only unknown is $\bar{q}$. Thus, after observing the test scores,the school should be able to accurately estimate $\bar{q}$ based on the observed distribution.

    This means that under Bayesian optimal estimation, the students who choose not to take the test will have the following estimated skill:

    $$
    \E\left[q|\{\theta_k\}_{k=1}^{K-1},Y=0\right]=\frac{\int_{-\infty}^{\bar{q}}\pi \text{Pr}\left(q=\pi| q\sim \mathcal{N}\left(\hat{\mu},\hat{\sigma}^2\right) \right)d\pi}{\int_{-\infty}^{\bar{q}}\text{Pr}\left(q=\pi| q\sim \mathcal{N}\left(\hat{\mu},\hat{\sigma}^2\right) \right)d\pi}\triangleq \tilde{q}.
    $$
    It is clear that $\tilde{q}<\bar{q}$. Moreover, we observe that by Bayesian optimal estimation, 
    $$
    \E\left[q|\{\theta_k\}_{k=1}^{K-1},Y=0\right]=\E\left[q|\{\theta_k\}_{k=1}^{K-1},Y=X=1, \theta_K=\tilde{q}\right].
    $$
    That is, $\tilde{q}$ happens to be the cutoff value $\theta'_K$ that we used in characterizing students behavior in equation \ref{eq:behavior_must_report}. Also, for $\tilde{q}\le q<\bar{q}$, we have $\text{Pr}(\theta_K>\tilde{q}|q)\ge \frac{1}{2}$. That is to say, there are students in this setting with skill level $\tilde{q}\le q<\bar{q}$ that are not currently taking the test, but can benefit from taking thus should take the test according to our behavior characterization, so this situation is not stable.

    Notice that the above argument holds whenever not all students take the test, so for fixed $\{\theta_k\}_{k=1}^{K-1}$, the unique equilibrium would be that all students take the test. Further extending to all possible $\{\theta_k\}_{k=1}^{K-1}$, we get that the unique equilibrium for all students is to take the test.

\end{proof}

\begin{customprop}{4.2}
Suppose the school observes access $Z$. Consider any policy $P$ that performs Bayesian optimal estimation on students with access. Then, $P$ is not latent skill fair. %
\end{customprop}

\begin{proof}
  Here it does not matter that strategy space the students have, since by Lemma \ref{lem:strategyaccess}, all students with access will take the test and report their score. Consider a set of students with fixed features $\{\theta_k\}_{k=1}^{K-1}$, and consider the following four subsets of them:
  \begin{itemize}
      \item Group A: $Z_A=0$, $q_A=q_1$
      \item Group B: $Z_B=0$, $q_B=q_2$
      \item Group C: $Z_C=1$, $q_C=q_1$
      \item Group D: $Z_D=1$, $q_D=q_2$
  \end{itemize}
  where we assume $q_1>q_2$. 
  
  First we note that, for the school, there is no way to distinguish students in groups $A$ and $B$: they all hand in the same information $\{\theta_k\}_{k=1}^{K-1}$, thus we have the estimated skill level should be distributed the same for these two groups:
  $$
  \tilde{q}|Z=0,\{\theta_k\}_{k=1}^{K-1},q_1,P \eqdist \tilde{q}|Z=0,\{\theta_k\}_{k=1}^{K-1},q_2,P
  $$
  
  On the other hand, we can calculate the distribution of the estimated skill level for groups $C$ and $D$:
  \begin{align*}
  &&\text{Group C: }\tilde{q}|Z=1,\{\theta_k\}_{k=1}^{K-1},q_1,P \sim \mathcal{N}\left(\frac{\mu\sigma^{-2}+\sum_{k=1}^{K-1}\theta_k\sigma_k^{-2}+q_1\sigma_K^{-2}}{\sigma^{-2}+\sum_{k=1}^{K}\sigma_k^{-2}},\frac{\sigma^{-2}+\sum_{k=1}^{K-1}\sigma_{k}^{-2}+2\sigma_K^{-2}}{\left(\sigma^{-2}+\sum_{k=1}^{K}\sigma_{k}^{-2}\right)^{2}}\right),\\
  &&\text{Group D: }\tilde{q}|Z=1,\{\theta_k\}_{k=1}^{K-1},q_2,P \sim \mathcal{N}\left(\frac{\mu\sigma^{-2}+\sum_{k=1}^{K-1}\theta_k\sigma_k^{-2}+q_2\sigma_K^{-2}}{\sigma^{-2}+\sum_{k=1}^{K}\sigma_k^{-2}},\frac{\sigma^{-2}+\sum_{k=1}^{K-1}\sigma_{k}^{-2}+2\sigma_K^{-2}}{\left(\sigma^{-2}+\sum_{k=1}^{K}\sigma_{k}^{-2}\right)^{2}}\right),
  \end{align*}
  and we immediately find that since $\sigma_K$ is positive, they are not equal in distribution:
  $$
  \tilde{q}|Z=1,\{\theta_k\}_{k=1}^{K-1},q_1,P\neqdist \tilde{q}|Z=1,\{\theta_k\}_{k=1}^{K-1},q_2,P.
  $$
  
  Thus trivially, at least one of the following must hold:
  $$
  \tilde{q}|Z=0,\{\theta_k\}_{k=1}^{K-1},q_1,P\neqdist \tilde{q}|Z=1,\{\theta_k\}_{k=1}^{K-1},q_1,P,
  $$
  or
  $$
  \tilde{q}|Z=0,\{\theta_k\}_{k=1}^{K-1},q_2,P\neqdist \tilde{q}|Z=1,\{\theta_k\}_{k=1}^{K-1},q_2,P.
  $$
  so latent skill fairness is not possible.
\end{proof}

\begin{customprop}{4.3}
    Suppose the school observes access status $Z$. Then, $P_{BO}$ -- the policy that uses Bayesian optimal estimation for all students -- is not observably or demographically fair.

\end{customprop}

\begin{proof}

From Lemma \ref{lem:strategyaccess}, all students with access will report the score. For a set of students with fixed features $\{\theta_k\}_{k=1}^{K-1}$, the distribution of their test score would be:
$$
\theta_K|\{\theta_k\}_{k=1}^{K-1}\sim \mathcal{N}\left(\frac{\mu\sigma^{-2}+\sum_{k=1}^{K-1}\theta_k\sigma_k^{-2}}{\sigma^{-2}+\sum_{i=1}^{K-1}\sigma_k^{-2}},\sigma_K^2+\frac{1}{\sigma^{-2}+\sum_{i=1}^{K-1}\sigma_k^{-2}}\right).
$$

Integrating over all possible test scores, we get that the expected estimated skill for these students would be:
\begin{align*}
\tilde{q}|Z=1,\{\theta_k\}_{k=1}^{K-1},P \sim \mathcal{N}\left(\frac{\mu\sigma^{-2}+\sum_{k=1}^{K-1}\theta_k\sigma_k^{-2}}{\sigma^{-2}+\sum_{k=1}^{K-1}\sigma_k^{-2}},\frac{\sigma_k^{-4}}{(\sigma^{-2}+\sum_{i=1}^{K}\sigma_k^{-2})^2}\left(\sigma_k^{2}+\frac{1}{\sigma^{-2}+\sum_{i=1}^{K-1}\sigma_k^{-2}}\right)\right).
\end{align*}

For the students without access to tests, their estimated skill is fixed under a Bayesian optimal estimation as:
\begin{align*}
    \tilde{q}|Z=0,\{\theta_k\}_{k=1}^{K-1},P = \frac{\mu\sigma^{-2}+\sum_{k=1}^{K-1}\theta_k\sigma_k^{-2}}{\sigma^{-2}+\sum_{k=1}^{K-1}\sigma_k^{-2}}.
\end{align*}

Therefore they are not equal in distribution, and observable fairness does not hold.

For the case with demographic fairness, note that students in both groups have their other features $\thetaset$ distributed exactly the same as assumed. Since $\tilde{q}|Z=0,P$ and $\tilde{q}|Z=1,P$ are the integration of $\tilde{q}|Z=0,\{\theta_k\}_{k=1}^{K-1},P$ and $\tilde{q}|Z=1,\{\theta_k\}_{k=1}^{K-1},P$ over all $\thetaset$, the inequality in distribution of the latter pair implies the inequality in distribution of the former, hence demographic fairness is impossible.

\end{proof}

\propfairnesssampling*

\begin{proof}
  Under the assumptions, following the proof of Proposition \ref{prop:knowobsfair}, the two quantities
  $$
  \tilde{q}|Z_i=0, \{\theta_k\}_{K=1}^{K-1}, P_{ME} \text{ and }\tilde{q}|Z_i=1, \{\theta_k\}_{K=1}^{K-1}, P_{ME}
  $$
  follow the same distribution, namely
  \begin{equation}
      \label{long_normal}
      \mathcal{N}\left(\frac{\mu\sigma^{-2}+\sum_{k=1}^{K-1}\theta_k\sigma_k^{-2}}{\sigma^{-2}+\sum_{k=1}^{K-1}\sigma_k^{-2}},\frac{\sigma_k^{-4}}{(\sigma^{-2}+\sum_{i=1}^{K}\sigma_k^{-2})^2}\left(\sigma_k^{2}+\frac{1}{\sigma^{-2}+\sum_{i=1}^{K-1}\sigma_k^{-2}}\right)\right)
  \end{equation}

So this policy is observably fair.    
\end{proof}

\end{document}